\documentclass[aps,prb,twocolumn,groupedaddress,showpacs]{revtex4-1}

\usepackage{graphicx,color,url,hyperref}

\begin{document}

\title{
Anisotropic ferromagnetism and structure stability in $4f$-$3d$ intermetallics:\\
{\it ab initio} structure optimization and magnetic anisotropy for RCo$_5$ (R=Ce, La, and Y)
}

\author{
Munehisa~Matsumoto
}
\affiliation{
Institute for Solid State Physics (ISSP), University of Tokyo, Kashiwa 277-8581, JAPAN
}

\date{\today}

\begin{abstract}
Electronic mechanism in the interplay between ferromagnetism and structure stability 
of $4f$-$3d$ intermetallics in the main phase of rare-earth permanent magnets
is investigated from first principles.
We present a case study with an archetypical materials family RCo$_5$ (R=Ce, La, Y),
which was a part of the earliest rare-earth permanent magnets and from which
other representative main-phase compounds can be regarded as a derived type.
Comparison with the champion magnet materials family R$_2$T$_{14}$B
and recently revisited materials family RT$_{12}$ (T=Co and Fe)
points to a direction
leading to a mid-class magnet for the possible next generation materials.
\end{abstract}


%

\maketitle

\section{Introduction}

Rare-earth permanent magnet (REPM) obviously
needs a good ferromagnet with sufficiently strong coercivity
and robust structure. Unfortunately,
cohesive energy and magnetic energy in magnetic materials
do not quite behave in a synchronized way:
too strong magnetization
may stretch the lattice spacing too much via the magnetovolume effect
for the chemical bonds to sustain the crystal structure. Thus a strong magnetization
can ruin the structure stability.
This makes one of the unavoidable trade-off situations
in the materials design for permanent magnets.
Careful inspection is needed to find out a best compromise
with an optimal chemical composition and crystal structure
to satisfy the prerequisites for permanent magnets. We do this
on the basis of {\it ab initio} electronic structure theory,
unearthing interplay in the middle
of dual nature harboring both of delocalization
and localization both in $3d$-electrons and in
$4f$-electrons.

We focus on light rare-earth elements, R=Ce and La, motivated by their abundance
and also in quest for a way
to possibly exploit the subtlety in
$4f$-electron physics to fabricate a new type of REPM's
where half-delocalized $4f$-electrons contribute positively to the bulk magnetic properties.
As a reference case without $4f$-electrons, R=Y is also addressed.
The particular crystal structure
of RCo$_5$ makes a part of the building block
for all of the other representative compounds, R$_2$M$_{14}$B~\cite{rmp_1991},
RM$_{12}$ and R$_2$M$_{17}$~\cite{liandcoey_1991} where M represents Fe-group elements.

In the next section, we describe
our methods based on {\it ab initio} structure optimization utilizing
the open-source package OpenMX~\cite{OpenMX}.
In Sec.~\ref{sec::results},
we present {\it ab initio} results on RCo$_5$ (R=Ce, La, and Y)
for their formation energy, magnetization, and magnetic anisotropy,
being contrasted to the analogous data for the Fe-counterparts.
Structure varieties and implications on the material-design
principles for REPM's are discussed in Sec.~\ref{sec::disc}.
Conclusions and outlook are given in the final section.

\section{Methods}
Intrinsic magnetic properties of RCo$_{5}$ and RFe$_5$
(R=Ce, La, Y)
are calculated from first principles via
{\it ab initio} structure optimization utilizing the open-source
software package OpenMX~\cite{OpenMX,Ozaki2003,Ozaki2004,Ozaki2005,Duy2014,Lejaeghere2016} on the basis of pseudopotentials~\cite{MBK1993,Theurich2001}
and the local orbital basis sets.

The lattice constants of Fe-group ferromagnets seems to be
best described within Generalized Gradient Approximation (GGA) as proposed
Perdew, Burke, and Ernzerhof (PBE)~\cite{prl_1996} and we present the results based on GGA-PBE.
The basis set we take in OpenMX is
\verb|Ce8.0-s2p2d2f1|,
\verb|La8.0-s2p2d2f1|,
\verb|Y8.0-s3p2d2f1|, \verb|Fe6.0S-s2p2d1|, and \verb|Co6.0S-s2p2d2f1| for R(Co,Fe)$_5$
within the given pseudopotential data set~\cite{OpenMX}.
The energy cutoff is set to be 500~Ry of which choice
has been inspected together with the basis sets to ensure a good convergence.

Similar basis sets with a few more or less inclusion of local basis wavefunctions
can be good as well depending on the target materials
and the issue being investigated as long as the choice of the basis set
is coherently applied in a fixed scope of target materials and target observables.
For the calculations of magnetic anisotropy energy presented below and elsewhere~\cite{shishido},
we have actually seen that slightly richer basis sets
\verb|Ce8.0-s3p3d3f2| and \verb|La8.0-s3p3d3f2| work on a par with
the basis sets written
in the previous paragraph or sometimes
in a better way especially when the target material
is close to the verge of a delocalization-localization transition in $4f$-electrons.
In the scope of the present work, presumably we stay on the side where the delocalized nature
of $4f$-electrons dominates within the crystal structure of CeCo$_5$.
For this purpose, either \verb|Ce8.0-s3p3d3f2| or \verb|Ce8.0-s2p2d2f1| will do basically.

The starting structure
is taken from the experimentally measured lattice constants for YCo$_5$~\cite{neutron_1980}
and {\it ab initio} structure optimization is done for stoichiometric compounds
RCo$_5$ and RFe$_5$
to get a minimized energy $U_{\rm tot}[\mbox{RT$_5$}]$
(T=Co, Fe) and the associated magnetization in the ground state. Thus extracted energy
is used to assess the structure stability by looking at the formation energy
referring to the elemental materials which are analogously addressed with {\it ab initio}
structure optimization by which the reference energies are extracted.
The formation energy of RCo$_5$, which we denote by $\Delta E[\mbox{RCo$_5$}]$,
is defined as follows:
\begin{eqnarray*}
&& \Delta E [\mbox{RCo$_5$}]\\
& \equiv & U_{\rm tot}[\mbox{RCo$_5$}]-U_{\rm tot}[\mbox{R per atom}]-5U_{\rm tot}[\mbox{Co per atom}]
\end{eqnarray*}
The structure optimization is done allowing for magnetic polarization without spin-orbit interaction.
Then on top of the optimized lattice,
magnetic anisotropy energy is investigated by fully relativistic calculations
incorporating the spin-orbit interaction, putting a constraint
on the direction of magnetization and numerically measuring the energy as a function of the angle
between magnetization and crystallographic $c$-axis.
Thus we look at the trends in the intrinsic properties focusing on the tradeoff
between formation energy and magnetization, assisted by the data for magnetic anisotropy.
As for the other prerequisite intrinsic property, Curie temperature,
some of the issues and finite-temperature magnetism are
addressed in separate works~\cite{mm_2014,mm_2015,mm_2018,chris_2018}.

\section{Results}
\label{sec::results}

{\it Ab initio} structure optimization for RCo$_5$, elemental R (fcc-Ce, dhcp-La, hcp-Y)
and elemental Co, that is, hcp-Co gives the energy
and magnetization for each of the target systems
within the given pseudopotential data sets.
Calculated formation energy and magnetization from structure optimization runs
are presented in Sec.~\ref{sec::RT5_deltaE}~and~\ref{sec::RT5_mag}, respectively.
Then in Sec.~\ref{sec::RT5_MAE} we show results from fully-relativistic calculations for magnetic anisotropy energy
on top of the optimized lattice.

\subsection{Formation energy}
\label{sec::RT5_deltaE}

\begin{figure}
\scalebox{0.7}{\includegraphics{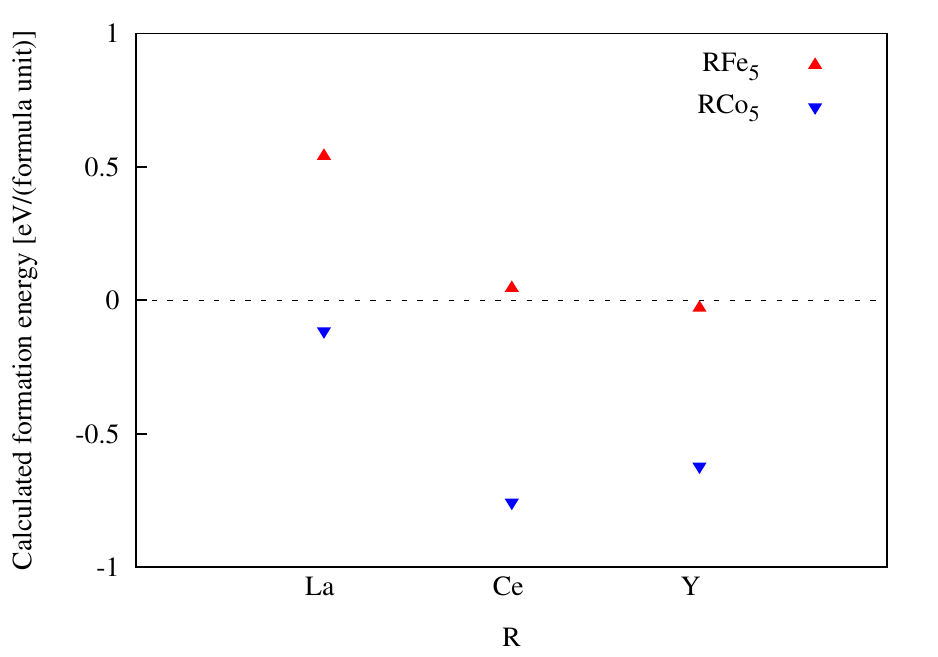}}
\caption{\label{fig::deltaE_RT5} Calculated formation energy for RT$_5$ (R=Ce, La, Y and T=Fe and Co).}
\end{figure}

We start with inspecting the formation energy as a clue
for the trend in the structure stability. Taking calculated energy
within the particular choice of the basis set described in the previous section
and given standard data sets of pseudopotentials~\cite{OpenMX},
calculated formation energy for RT$_5$ (R=Y, La, Ce, and T=Co, Fe) is
summarized in Fig.~\ref{fig::deltaE_RT5}. The detailed procedures
for the calculation of formation energy follow those described in
Refs.~\onlinecite{mm_20181228}~and~\onlinecite{mm_20190129}.
On the optimized lattice, the lattice constants and unit-cell volume
have been read off as summarized in Table~\ref{table::latt}.
Comparing with the experimental data~\cite{theory_1990}
which is available only for Co-based materials,
it is seen that {\it ab initio} structure optimization predicts
the realistic lattice constants and the unit cell volume within the precision
of three significant digits.

We reproduce the known experimental fact that RFe$_5$ is metastable~\cite{metastable_review}
with the calculated formation energy for RFe$_5$ running into the  positive region.
The relative trend between LaCo$_5$ and YCo$_5$,
\[
\left|\Delta E[\mbox{LaCo$_5$}]\right| < \left|\Delta E[\mbox{YCo$_5$}]\right|
\]
shows that the smaller rare-earth elements in the overall trend of lanthanide contraction
comes with the better structure stability. 
Most notable difference between LaCo$_5$ and YCo$_5$ lies
in the radius of rare-earth element, with La being the largest rare earth element
and Y being as small as typical heavy rare earth elements like Dy in the trend of lanthanide contraction.
It is reasonable for La-based
compounds having the enlarged lattice with La sitting on the peak in the size of rare-earth atoms
to become relatively fragile against the anticipated magnetovolume effect when combined
with the strong $3d$-electron magnetization coming from Fe-group elements,
since there would be relatively
less margin for the chemical bonds to hold on against
the volume expansion as imposed by magnetization.
In this regard, the relatively small size of Ce$^{4+}$ actually gives an
advantage on Ce-based ferromagnets at least for the structure stability.
\begin{table}
\begin{tabular}{ccccc} \hline 
 & \multicolumn{2}{c}{calculated results}
 & \multicolumn{2}{c}{experimental data} \\ \hline
 & $(a,c)~\mbox{[\AA]}$ & $V_{\rm cell}$~[\AA$^3$] & $(a,c)~\mbox{[\AA]}$ & $V_{\rm cell}$~[\AA$^3$] \\  \hline
CeCo$_5$ & $(4.89,4.02)$ & $84.0$ & $(4.93, 4.02)$ & $84.5$ \\  \hline
LaCo$_5$ & $(5.06,3.96)$ & $87.7$ & $(5.09, 3.94)$ & $88.3$ \\  \hline
 YCo$_5$ & $(4.92,3.95)$ & $83.2$ & $(4.93, 3.99)$ & $84.0$ \\  \hline\hline
CeFe$_5$ & $(5.06,4.10)$ & $91.1$ & \multicolumn{2}{c}{N/A} \\  \hline
LaFe$_5$ & $(5.19,4.08)$ & $95.0$ & \multicolumn{2}{c}{N/A} \\  \hline
 YFe$_5$ & $(5.08,3.99)$ & $88.8$ & \multicolumn{2}{c}{N/A} \\  \hline
\end{tabular}
\caption{\label{table::latt} Optimized lattice constants
and unit-cell volume $V_{\rm cell}$ for RT$_5$ (R=Ce, La, Y and T=Co, Fe).
The experimental data are taken and rounded up to the 3rd digit
as quoted in Ref.~\onlinecite{theory_1990}.}
\end{table}

\subsection{Magnetization}
\label{sec::RT5_mag}

Calculated magnetization as a result of the structure optimization is
summarized in Table~\ref{table::mag}. Total magnetic moments per formula unit
which occupies the unit cell
is normalized by the volume of the unit cell to yield the magnetization measured in Tesla
which is of direct relevance for REPM's. The quantitative agreement between
the results from {\it ab initio} structure optimization and
experiments for the magnetization in Tesla is seen up to two digits.
The realistic energy scales for the cohesion and magnetism seem
to be properly taken into account in the present description.
\begin{table}
\begin{tabular}{cccc} \hline 
 & \multicolumn{2}{c}{calculated results} & expt. \\  \hline
 & $M~\mbox{[$\mu_{\rm B}$/(f.u.)]}$ & $M$~[Tesla] & $M$~[Tesla] \\  \hline
CeCo$_5$ & $6.03$ & $0.837$ & $0.77$\\  \hline
LaCo$_5$ & $7.03$ & $0.934$ & $0.91$\\  \hline
 YCo$_5$ & $7.22$ & $1.01$  & $1.09$ \\  \hline\hline
CeFe$_5$ & $10.6$ & $1.36$  & N/A \\  \hline
LaFe$_5$ & $11.7$ & $1.43$  & N/A \\  \hline
 YFe$_5$ & $10.7$ & $1.41$  & N/A \\  \hline
\end{tabular}
\caption{\label{table::mag} Calculated magnetization for RT$_5$ (R=Ce, La, Y, and T=Co, Fe).
Experimental data for magnetization in Tesla is taken from Ref.~\onlinecite{ohashi_2012}.}
\end{table}

\begin{figure}
\begin{tabular}{l}
(a) \\
\scalebox{0.7}{\includegraphics{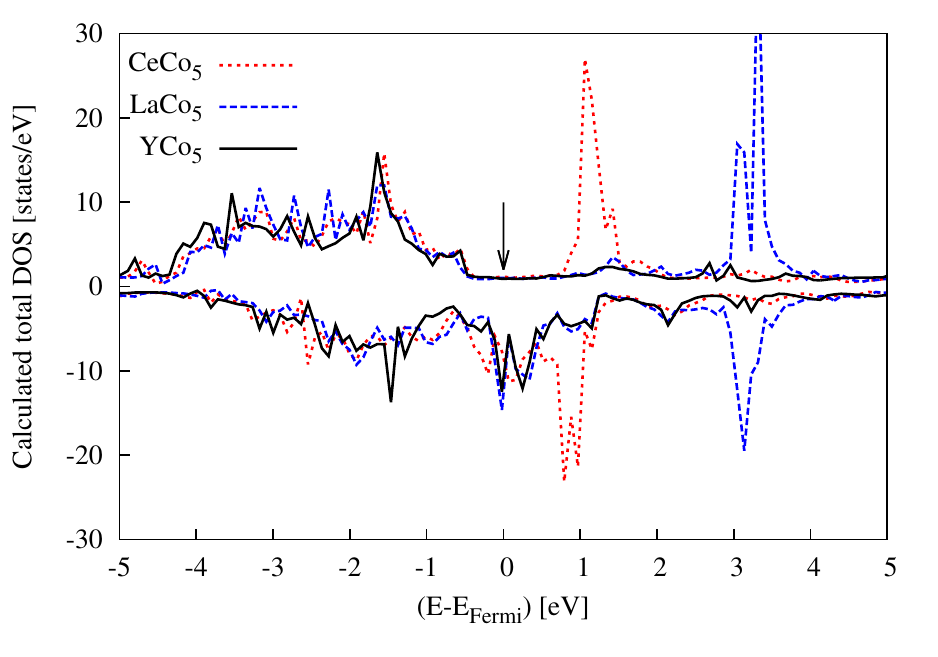}} \\
(b) \\
\scalebox{0.7}{\includegraphics{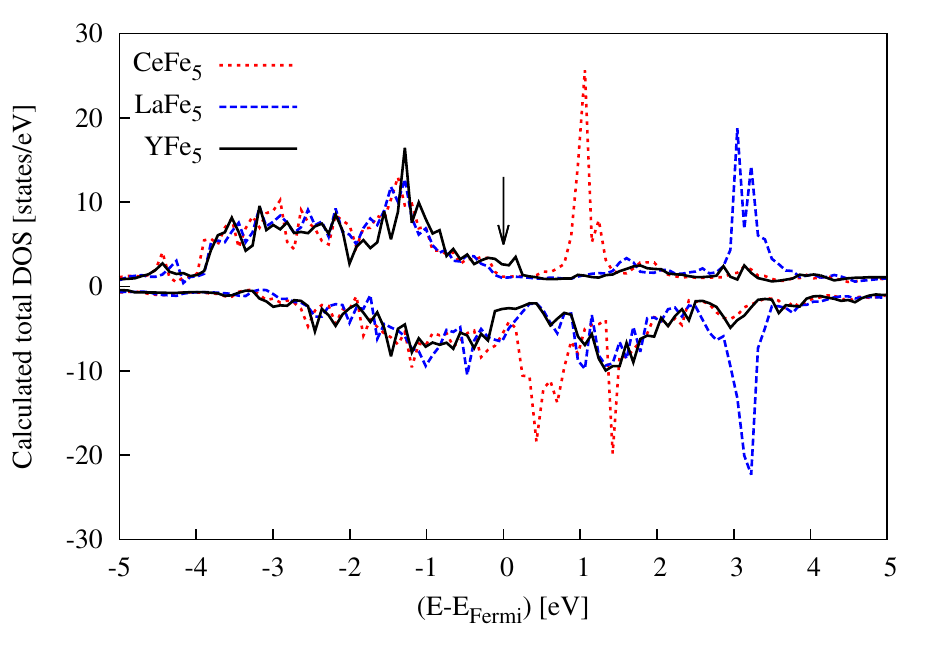}}
\end{tabular}
\caption{\label{fig::DOS_RT5} (Color online)
Calculated total
density of states for (a) RCo$_5$ and (b) RFe$_5$ (R=Ce, La, and Y).
The arrow in the figure points to the majority-spin states on the Fermi level.}
\end{figure}

Nature of ferromagnetism
can be inspected from 
calculated total density of states (DOS)
as shown in Fig.~\ref{fig::DOS_RT5}.
Only for the fictitious YFe$_5$, weak ferromagnetism with a non-negligible amount
of majority-spin states on the Fermi level is seen. All of the other compounds
show strong ferromagnetism with the $d$-electron majority-spin states all below the Fermi level
which means that ferromagnetic order is basically saturated there and there is little space
for further enhancing magnetization e.g. by mixing Fe and Co to follow the celebrated Slater-Pauling curve.
We note that our DOS for CeCo$_5$ look at variance with what is shown
in one of the previous works~\cite{theory_1990} where their DOS could
point to the possible weak ferromagnetism in CeCo$_5$. We have actually verified
in our calculations that Slater-Pauling curve is not observed for Ce(Co,Fe)$_5$~\cite{with_martin}
so indeed weak ferromagnetism does not seems to be the case at least for the optimized CeCo$_5$
in our calculations. Since the electronic states near the Fermi level can sensitively
depend on the details of calculation setup it could be reasonable to get some variant results
for CeCo$_5$ that may reside almost on a border line between weak ferromagnetism and strong ferromagnetism.

\subsection{
Magnetic anisotropy energy (MAE)
}
\label{sec::RT5_MAE}

\begin{figure}
\scalebox{0.7}{\includegraphics{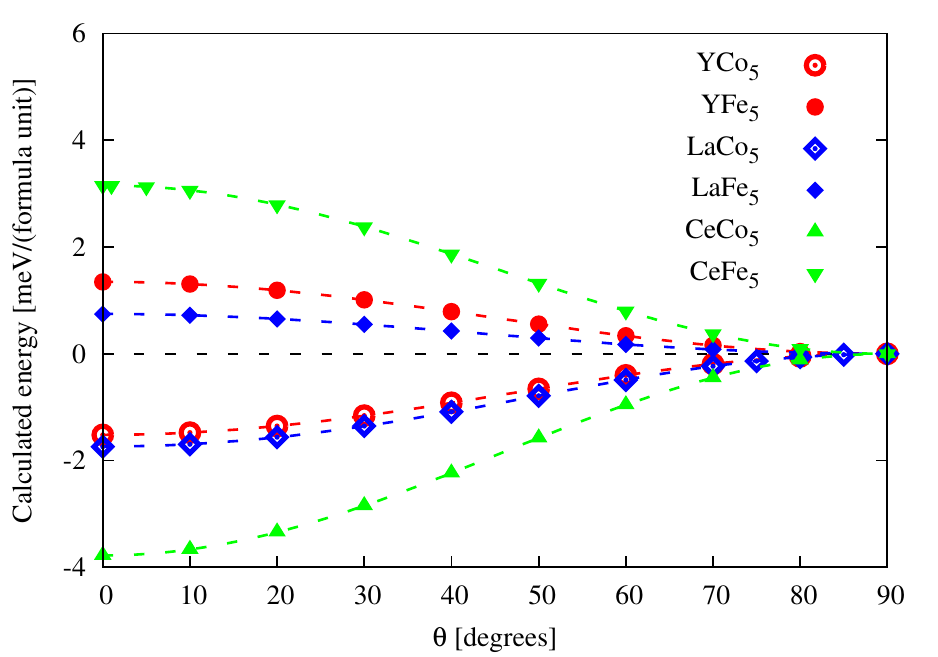}}
\caption{\label{fig::rotation_runs} (Color online) angle dependence of the calculated energy
with the zero point being set at the bottom of the calculated total energy. Here the angle
is made between the bulk magnetization direction and the crystallographic
$c$-axis. Overall RCo$_5$ show uni-axial magnetic anisotropy while RFe$_5$'s have
easy-direction along the $ab$-plane. Dotted lines are fits of Eq.~(\ref{eq::aniso}) in the text
to the calculated data points. The results of such fits are summarized in Table~\ref{table::mae}.}
\end{figure}

On top of the optimized crystal structure, magnetic anisotropy energy
is inspected from first principles by monitoring the energy as a function of the angle
between constrained magnetization and the crystallographic $c$-axis
under the presence of spin-orbit interaction. The results
are shown in Fig.~\ref{fig::rotation_runs}.
Uni-axial magnetic anisotropy of RCo$_5$
is switched into easy-plane magnetic anisotropy with RFe$_5$.
\begin{table}
\begin{tabular}{cccc} \hline 
 & $K~\mbox{[meV/(formula unit)]}$ & $p$ & $q$\\  \hline
CeCo$_5$ & $3.79(2)$ & $-0.0181(2)$ & $0.004(4)$ \\  \hline
LaCo$_5$ & $1.737(1)$ & $-0.162(4)$ & $q\equiv 0$ \\  \hline
YCo$_5$ & $1.5171(6)$ & $-0.077(2)$ & $q\equiv 0$\\  \hline\hline
CeFe$_5$ & $-4.39(6)$ & $-0.0130(3)$ & $0.28(1)$  \\  \hline
LaFe$_5$ & $-0.7471(7)$ & $0.073(4)$ & $q\equiv 0$ \\  \hline
YFe$_5$ & $-1.34706(3)$ & $-0.00103(8)$ & $q\equiv 0$ \\  \hline
\end{tabular}
\caption{\label{table::mae} Calculated magnetic anisotropy energy and the coefficient
of the higher order terms for RT$_5$ (R=Ce, La, Y and T= Co, Fe).}
\end{table}
Calculated energy as a function of the magnetization direction as shown in Fig.~\ref{fig::rotation_runs}
is fitted to the following relation
\begin{equation}
E_{\rm aniso}=-K\left[ (1-p-q)\cos^2\theta + p \cos^4\theta + q \cos^6\theta \right]
\label{eq::aniso}
\end{equation}
to extract the magnetic anisotropy energy together with the higher-order terms. The fit results
are summarized in Table~\ref{table::mae}. The absolute values of $K$
show a significant underestimate as compared to experiments,
which seems to be reasonable for {\it ab initio} calculations for $3d$-electron
magnetic anisotropy.
Without special treatment such as orbital polarization,
LDA/GGA+U, or self-interaction correction,
it is not easy to quantitatively match the calculated MAE
to the experimentally observed range~\cite{mazin_2003}.

For the moment we focus on qualitative trends in the extracted parameters.
It is to be noted for the $4f$-electron part
that relatively strong anisotropy is found for Ce-based compounds 
as compared to La or Y-based compounds even though the $4f$-electrons
are presumably in a delocalized state. In Fig.~\ref{fig::DOS_RT5}
the majority-spin state for Ce, seen on the opposite side
to the majority-spin of Co or Fe, has a significant contribution
above and below the Fermi level which means $4f$-electrons are half-localized
as are $3d$-electrons in Fe and Co. Thus they contribute positively to the bulk anisotropy.

Turning to the $3d$-electron part,
it is seen that
all Fe-based compounds end up with easy-plane anisotropy: this is not quite
interesting in the context of REPM. Given that Fe-rich compounds come with
poor structure stability and loss of uni-axial magnetic anisotropy,
relatively strong magnetization does not help by itself and an optimal chemical
composition for REPM would rather lie on the Co-rich side. Quantitatively
identifying the location of an optimal point on a continuum of
chemical composition axis is addressed in a separate work~\cite{with_martin}.

Remarkably, the coefficient of the extracted higher-order coefficient in YCo$_5$ 
is in agreement with the past experimental claim stating $K_2/K_1=-0.021$~\cite{alameda_1981}
where $K_1\equiv K(1-p)$ and $K_2\equiv Kp$ in our notation. Our data amounts
to $K_2/K_1=-0.071$ and the sign of the higher order contribution has been correctly reproduced.

\section{Discussions}
\label{sec::disc}

Comparison between
what we have seen for RT$_5$ and analogous data for the champion magnet compound
family, R$_2$T$_{14}$B, is described in Sec.~\ref{sec::RT5_vs_R2T14B}
which reveals the trends with respect to the crystal structure variety
where the latter was reached in quest for a way to enlarge the interatomic spacing
between Fe~\cite{sagawa_1984}. Implications on the
materials design for REPM's are discussed in Sec.~\ref{sec::RT5_vs_RT12}
taking the recently discussed materials family RT$_{12}$~\cite{JOM} as the possible
playground for the application of the proposed principles.

\subsection{Comparison with R$_2$T$_{14}$B}
\label{sec::RT5_vs_R2T14B}

\subsubsection{Formation energy}
\begin{figure}
\scalebox{0.7}{\includegraphics{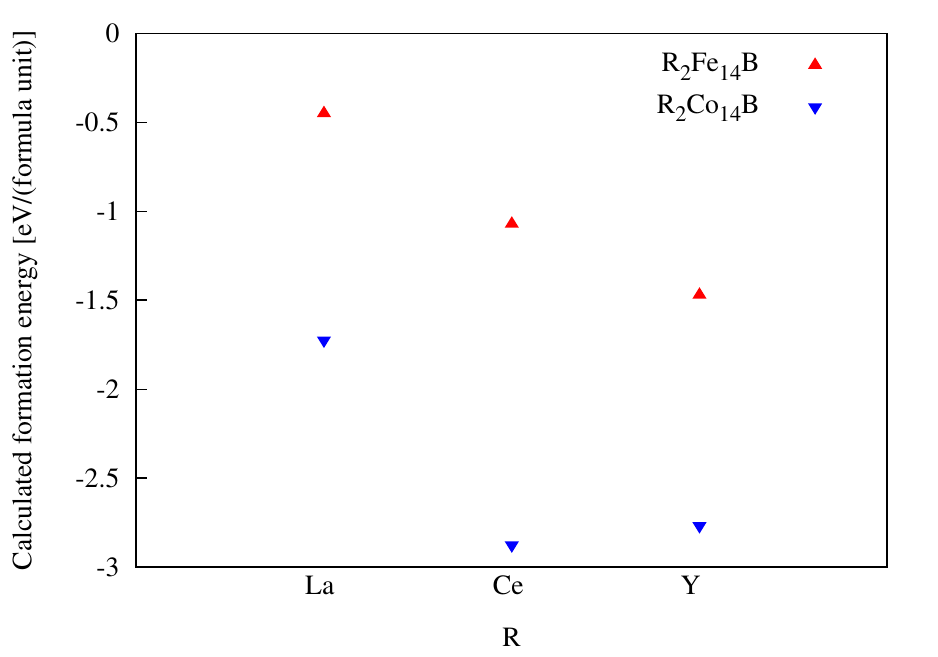}} 
\caption{\label{fig::deltaE_R2T14B} Calculated formation energy for
R$_2$T$_{14}$B (R=Ce, La, Y and T=Fe,Co). The data
for (La,Ce)$_2$Fe$_{14}$B and Y$_2$(Fe,Co)$_{14}$B are taken
from Ref.~\onlinecite{mm_20190129} and Ref.~\onlinecite{mm_20181228}, respectively.}
\end{figure}

Calculated formation energy for Ce$_2$Co$_{14}$B and La$_2$Co$_{14}$B are shown
in Fig.~\ref{fig::deltaE_R2T14B} together with the data
for R$_2$Fe$_{14}$B (R=Ce and La)~\cite{mm_20190129}
and Y$_2$T$_{14}$B (T=Co, Fe)~\cite{mm_20181228} for the convenience of getting an overview.
Parallel trends with what is seen in Fig.~\ref{fig::deltaE_RT5} is obvious, with the most significant
difference being the overall offset to push down everything down to the negative region of the formation energy.
Even with the intrinsic trend that Fe-based compounds come with the relatively poor structure stability,
with the particular crystal structure of R$_2$T$_{14}$B they can be stabilized except for La$_2$Fe$_{14}$B
that looks perhaps too close to being on the verge to the positive side of the formation energy.

Absence of Ce$_2$Co$_{14}$B in the experimental
literature is a mystery as long as we look at the trends
of the formation energy. The absolute value of the calculated formation energy for Ce$_2$Co$_{14}$B
is more than double of that for Ce$_2$Fe$_{14}$B, which is consistent with one of the recent claims~\cite{herbst_2017}.
Presumably, other compounds may compete against Ce$_2$Co$_{14}$B in getting bulk-stabilized, or something beyond
the present level of electronic structure calculations might be at work in real experiments:
in the present calculations we have plainly put $4f$-electrons of Ce into the valence state
which is not always justified.

\subsubsection{Magnetization}
Calculated magnetization for R$_2$T$_{14}$B are summarized
in Table~\ref{table::mag_and_cell_R2T14B}~and~\ref{table::mag_R2T14B}
together with the optimized lattice information for the convenience of calculating
magnetization in Tesla. Again the data for R$_2$Fe$_{14}$B (R=Ce and La)
and Y$_2$T$_{14}$B (T=Co, Fe) are taken from Refs.~\onlinecite{mm_20190129}~and~\onlinecite{mm_20181228}, respectively.
It is seen that the optimized lattice constants are in quantitative agreement with the experimental ones
up to three digits while calculated magnetization in Tesla sometimes has an overestimate
where the deviation from experimental data amounts to a few
tens of \%. Practically it has not been
easy to let the calculations for R$_2$Co$_{14}$B converge. It actually took
having an initial magnetic state with each Co atom starting with
unrealistically large magnetic polarization to reach any sensibly
converged results in our {\it ab initio} structure optimization for R$_2$Co$_{14}$B so far.
In this way it might have been reasonable to have ended up with an overestimate
of magnetization and also this may indicate that our description of $3d$-electron state may not have been
as precise as has been achieved for Fe-based counterparts~\cite{mm_20190129}.
Anyways relative trends e.g. between La$_2$Co$_{14}$B and Y$_2$Co$_{14}$B
seems to have been satisfactorily described.
\begin{table}
\begin{tabular}{ccccc} \hline
 & \multicolumn{3}{c}{calculated results} & expt. \\ \hline
 & $M~\mbox{[$\mu_B$/(f.u.)]}$  & $V_{\rm cell}~\mbox{[\AA$^3$]}$ & $(a,c)$~\AA  & $(a,c)$~\AA \\  \hline
Ce$_2$Co$_{14}$B & $16.64$ & $865$ & $(8.596,11.73)$ & N/A \\  \hline
La$_2$Co$_{14}$B & $18.67$ & $891$ & $(8.635,11.96)$ & $(8.67, 12.01)$ \\  \hline\hline
Y$_2$Co$_{14}$B & $18.70$ & $855$ & $(8.560,11.68)$ & $(8.60, 11.71)$ \\  \hline
Ce$_2$Fe$_{14}$B & $29.95$ & $936$ & $(8.797,12.11)$ & $(8.76, 12.11)$ \\  \hline
La$_2$Fe$_{14}$B & $31.92$ & $961$ & $(8.835,12.33)$ & $(8.82, 12.34)$ \\  \hline
Y$_2$Fe$_{14}$B & $31.47$ & $922$ & $(8.775,11.99)$ & $(8.76, 12.00)$ \\  \hline
\end{tabular}
\caption{\label{table::mag_and_cell_R2T14B}
Optimized lattice constants and unit-cell volume 
for R$_2$T$_{14}$B (R=Ce, La, Y
and T=Co, Fe). The data
for (La,Ce)$_2$Fe$_{14}$B and Y$_2$(Fe,Co)$_{14}$B
are partly
taken from Ref.~\onlinecite{mm_20190129} and Ref.~\onlinecite{mm_20181228}, respectively.
Experimental lattice constants are taken from Ref.~\onlinecite{rmp_1991}.
}
\end{table}

\begin{table}
\begin{tabular}{ccc} \hline
 & calc. & expt.\\  \hline
Ce$_2$Co$_{14}$B & $0.896$ & N/A \\  \hline
La$_2$Co$_{14}$B & $0.977$ & $0.741$ \\  \hline\hline
Y$_2$Co$_{14}$B & $1.02$ & $0.848$ \\  \hline
Ce$_2$Fe$_{14}$B & $1.49$ & $1.47$ \\  \hline
La$_2$Fe$_{14}$B & $1.55$ & $1.48$ \\  \hline
Y$_2$Fe$_{14}$B & $1.59$ & $1.59$ \\  \hline
\end{tabular}
\caption{\label{table::mag_R2T14B} Calculated magnetization in Tesla
for R$_2$T$_{14}$B (R=Ce, La, Y
and T=Co, Fe). Experimental data for $M$ at $T=4.2~\mbox{K}$
is taken from Ref.~\onlinecite{rmp_1991} for R$_2$Fe$_{14}$B
and Ref.~\onlinecite{buschow_1988} for R$_2$Co$_{14}$B.}
\end{table}

\subsubsection{Magnetic anisotropy energy}
\begin{figure}
\scalebox{0.7}{
\includegraphics{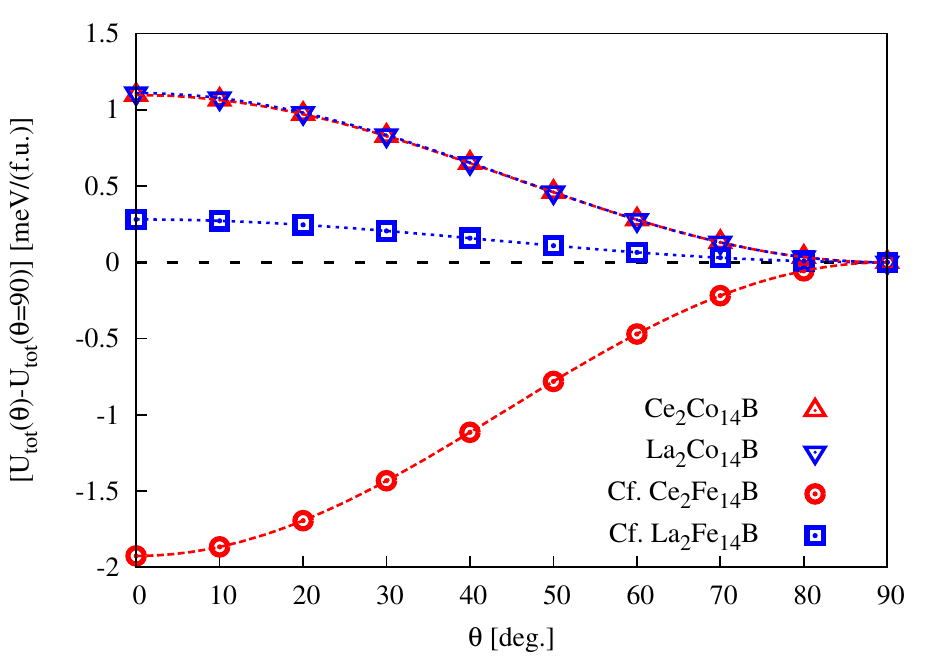}}
\caption{\label{fig::mae_R2T14B}
Angle-dependence of calculated energy of R$_{2}$Co$_{14}$B (R=Ce and La).
For a comparison, analogous data for R$_{2}$Fe$_{14}$B are included
partly from Ref.~\onlinecite{mm_20190129}.}
\end{figure}

Magnetic anisotropy energy for Ce$_2$Co$_{14}$B and La$_2$Co$_{14}$B
can be inspected in an analogous way to what was done for RT$_5$ in Sec.~\ref{sec::RT5_MAE}.
The results are shown in Fig.~\ref{fig::mae_R2T14B} together with the counterpart data
for Ce$_2$Fe$_{14}$B and La$_2$Fe$_{14}$B as partly taken from Ref.~\onlinecite{mm_20190129}.
It actually took lifting off the constraints on magnetic moments on Ce and La
for the fully relativistic runs of R$_2$Co$_{14}$B (R=Ce and La) to reach convergence.
The same was true for La$_2$Fe$_{14}$B in Ref.~\onlinecite{mm_20190129} but
it was not actually the case for Ce$_2$Fe$_{14}$B in Ref.~\onlinecite{mm_20190129} -
here, just for a comparison on an equal footing, results from
fully relativistic calculations for Ce$_2$Fe$_{14}$B without the constraints on magnetic moments on Ce are
presented. The difference from the fully relativistic calculations for Ce$_2$Fe$_{14}$B
in Ref.~\onlinecite{mm_20190129} where the constraints on the direction of magnetic moments were
all imposed on Ce and Fe is seen only quantitatively
in the absolute value of $K$ with a slightly smaller number
coming up here. The other parameters show qualitatively similar behavior, with
positive $p$ and negative $q$ coming up in the same range in the absolute value $|p|\simeq |q|$
for Ce$_2$Fe$_{14}$B, irrespectively of the constraints on the magnetic moments on Ce
being applied (Ref.~\onlinecite{mm_20190129}) or not (here). 
\begin{table}
\begin{tabular}{cccc} \hline 
 & $K~\mbox{[meV/(formula unit)]}$ & $p$ & $q$\\  \hline
Ce$_2$Co$_{14}$B & $-1.21(1)$ & $-0.0188(4)$ & $0.089(9)$ \\  \hline
La$_2$Co$_{14}$B & $-1.10974(5)$ & $-0.0072(2)$ & $q\equiv 0$ \\  \hline\hline
Ce$_2$Fe$_{14}$B & $1.83(1)$ & $0.0360(6)$ & $-0.051(7)$  \\  \hline
La$_2$Fe$_{14}$B & $-0.28106(7)$ & $0.098(1)$ & $q\equiv 0$ \\  \hline
\end{tabular}
\caption{\label{table::mae_R2T14B}
Calculated magnetic anisotropy energy and the coefficient of the higher order terms
for R$_2$T$_{14}$B.}
\end{table}

It is remarkable that we hardly observe
any difference between Ce$_2$Co$_{14}$B and La$_2$Co$_{14}$B
concerning their easy-plane bulk MAE in the absolute value, except for the qualitative behavior
of higher-order terms. This is in strong contrast to the difference between
Ce$_2$Fe$_{14}$B and La$_2$Fe$_{14}$B where even the sign of bulk anisotropy
is flipped between them.

\subsection{Implications on the possible new REPM based on RT$_{12}$}
\label{sec::RT5_vs_RT12}
In the past five years there has been a surge in the interests in RT$_{12}$ compounds~\cite{JOM}
that can potentially go beyond the champion magnet compounds R$_2$T$_{14}$B
in the intrinsic properties. Here the structure stability has been traded off
to demonstrate the real-material implementation of case studies
to go beyond Nd$_2$Fe$_{14}$B~\cite{miyake_2014,hirayama_2015,hirayama_2017}.
Recent interests focus on the quest for gaining both of
satisfactorily good magnetic properties and sufficient structure stability
to move on to a working REPM in industrial applications.
Here we take a brief look at the outcome of {\it ab initio} structure optimization
for RT$_{12}$ with R=Ce, La, Y and T=Co, Fe and see to what extent we might be able
to recycle what we have seen with RT$_5$ and R$_2$T$_{14}$B.

\begin{figure}
\scalebox{0.7}{\includegraphics{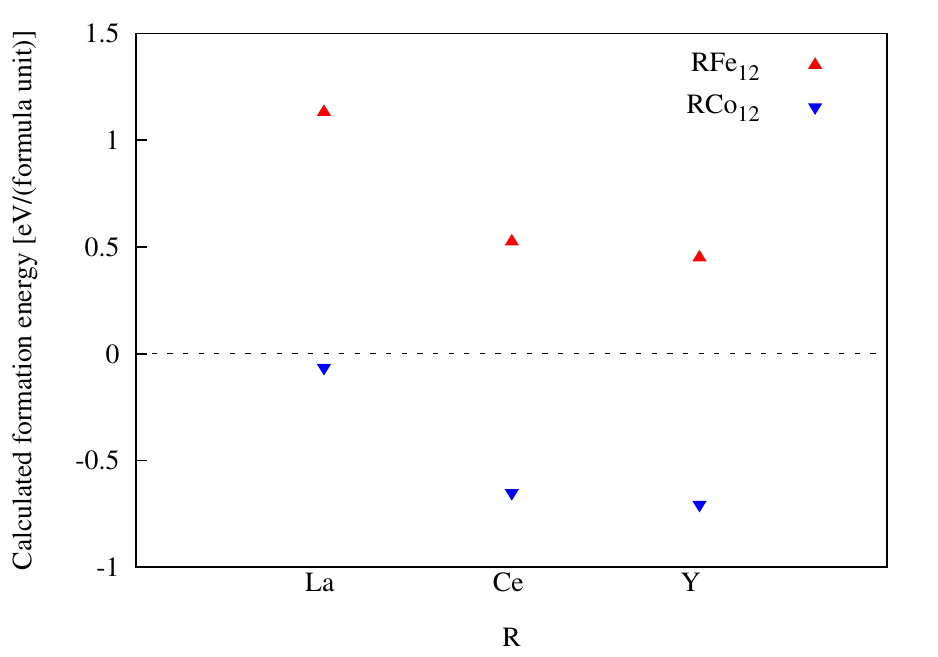}} \\
\caption{\label{fig::deltaE_RT12} Calculated formation energy for RT$_{12}$  (R=Ce, La, Y and T=Co, Fe).}
\end{figure}
\begin{table}
\begin{tabular}{ccccc} \hline
& \multicolumn{4}{c}{calculated results} \\ \hline
& $M$~[$\mu_{\rm B}$/(f.u.)] & $(a,c)$~\AA & $V_{\rm cell}$ & $M$~[Tesla] \\  \hline
CeCo$_{12}$ & $17.99$ & $(8.33, 4.65)$ & $161$ & $1.30$ \\ \hline
LaCo$_{12}$ & $18.95$ & $(8.40, 4.67)$ & $165$ & $1.34$ \\  \hline
YCo$_{12}$ & $19.05$ & $(8.28, 4.66)$ & $160$ & $1.39$ \\  \hline\hline
CeFe$_{12}$ & $27.15$ & $(8.58, 4.73)$ & $174$ & $1.82$ \\  \hline
LaFe$_{12}$ & $27.7$ & $(8.66, 4.73)$ & $177$ & $1.82$ \\  \hline
YFe$_{12}$ & $27.2$ & $(8.51,4.73)$ & $171$ & $1.85$ \\  \hline
\end{tabular}
\caption{\label{table::mag_RT12} Calculated magnetization and optimized lattice constants
and unit-cell volume
for RT$_{12}$ (R=Ce, La, Y and T=Co, Fe).}
\end{table}
Calculated formation energy for RT$_{12}$ is
shown in Fig.~\ref{fig::deltaE_RT12}.
The overall trend in the structure stability is qualitatively
shared among RT$_5$ and R$_2$T$_{14}$B in the sense
that a) Co-based compounds
are generally more stable than Fe-based compounds
and b) smaller rare-earth elements give more stable structure.
Combined with calculated magnetization in Tesla as summarized in Table~\ref{table::mag_RT12}
points to CeCo$_{12}$ and YCo$_{12}$ that can provide a good compromise coming with
both reasonably good magnetization and comfortably as stable as CeCo$_5$ as long as the calculated
formation energy shows. Apparently RT$_{12}$ resembles RT$_5$ in several respects
rather than being considered as an alternative to R$_2$T$_{14}$B. It might be feasible
to pursue the possible line toward a mid-class REPM on the basis of CeCo$_{12}$
possibly with a small addition of Fe for gaining magnetization not to the extent
where the structure stability is ruined,
in an analogous way to the development of Fe-doped YCo$_5$~\cite{chris_2018,marko,with_martin}.

\section{Conclusions and outlook}
\label{sec::conclusions}

Intrinsic trade-off between structure stability and magnetization
has been inspected from first principles for RT$_5$. Comparison has been
done with the champion magnet compounds R$_2$T$_{14}$B and the recently
focused materials family RT$_{12}$. Detailed characterization of the fictitious compound
Ce$_2$Co$_{14}$B has been done for the first time to the best of the author's knowledge.
Throughout the calculations covering materials family,
roles of Ce in helping structure stabilization
and positive contribution to the intrinsic magnetic anisotropy has been elucidated
on the basis of calculated electronic structure and we identify CeCo$_{12}$ as a possible starting
point for the fabrication of a next-generation mid-range REPM.

\begin{acknowledgments}
MM's work in ISSP, Univ.~of~Tokyo is supported by Toyota~Motor~Corporation.
Helpful interactions with A.~Marmodoro, A.~Ernst,
C.~E.~Patrick, J.~B.~Staunton,
M.~Hoffmann, H.~Akai,
T.~Miyake, Y.~Harashima
in related projects and useful comments given by
T.~Ozaki, F.~Ishii for {\it ab initio} calculations employing OpenMX are gratefully acknowledged.
Part of the present work was supported by
JSPS KAKENHI Grant No.~15K13525
and the Elements Strategy Initiative Center for Magnetic Materials
(ESICMM) under the outsourcing project of the Ministry of Education,
Culture, Sports, Science and Technology (MEXT), Japan.
Numerical calculations were done on ISSP supercomputer center, Univ. of Tokyo.
\end{acknowledgments}

\end{document}